# QoS Survey in IPv6 and Queuing Methods

Biju Issac
PhD Scholar (Part-time), Faculty of Engineering,
University Malaysia Sarawak, Malaysia.
[Lecturer, Swinburne University of Tech, Sarawak]
Tel : +60-82-416353, Fax : +60-82-423594
bissac@swinburne.edu.my

Khairuddhin Ab Hamid and C. E. Tan
Faculty of Engineering/Faculty of Computer
Science and Information Technology
University Malaysia Sarawak,
94300 Kota Samarahan, Sarawak, Malaysia.
khair@cans.unimas.my, cetan@fit.unimas.my

*Abstract* – **The routes in IP networks are determined by the IP destination address and the routing tables in each router on the path to the destination. Hence all the IP packets follow the same route until the route is changes due to congestion, link failure or topology updates. IPv4 tried using Type of Service (TOS) field in the IP header to classify traffic and that did not succeed as it was based on fair self-classification of applications in comparison to the network traffic of other applications. As multimedia applications were quite foreign at the initial IPv4 stage, TOS field was not used uniformly. As there are different existing Quality of Service (QoS) paradigms available, IPv6 QoS approach was designed to be more flexible. The IPv6 protocol thus has QoS-specific elements in Base header and Extension headers which can be used in different ways to enhance multimedia application performance. In this paper, we plan to survey these options and other QoS architectures and discuss their strengths and weaknesses. Some basic simulation for various queuing schemes is presented for comparison and a new queuing scheme prioritized WFQ with RR is proposed.**

## I. INTRODUCTION

There are different Quality of Service (QoS) paradigms especially in identifying the right location for providing QoS in a network. *The End-System based QoS* is simple but does not scale well for multimedia intensive applications. It takes the position that IP networks have no knowledge of network traffic characteristics and all traffic is transmitted on best effort basis. Since any intelligent traffic handling is absent, the end systems should take care of delay and other network spikes or jitters by themselves. The *Service-based QoS* in IPv6 defines different service classes. This is implemented by different multicast groups, which would be handled accordingly through different queuing and processing by the routers and end systems. For example, three classes of same audio or video traffic, each encoded with different quality can be defined. Here the traffic characteristics can be shaped at the source and intermediate routers through the use of IPv6 ICMP multicast feedback control message scheme. The *Class-based QoS* in IPv6 allows the routers to have clear information on how to deal with packets with different service requirements, through signaling information and other details in the IP packets. It uses IPv6 Routing Extension header for special route setup and Hop-by-Hop Extension header to transmit control information to all routers in the path. Here the network uses best-effort mode and it may implement a feedback to the loop to the sender to correct and adapt to the QoS requirements. For example, hierarchical coding may be used to transmit audio or video messages in three different encoding formats. Routers can drop packets if there is congestion say, starting from third to first. Even though the packet quality may differ in the three encodings, the packet is completely lost only if all the three encoded packets are dropped. Obviously, the additional overhead on the sender is a negative point, along with the non-guaranteed service quality on the receiver side. The *Resource Reservation-based QoS* requires full knowledge of connection details and QoS needs to reserve enough resources, for packet processing. This is a complex task compared to the other approaches. It requires the source and destination nodes to exchange sufficient signaling and control information, say using Resource Reservation Protocol (RSVP). It establishes packet classification and forward status information in each router along the path, thus introducing state information in a previously stateless IP datagram forwarding network. With IPv6 the implementation of this is much easier [1].

IPv4 networks has lot of problems with QoS like – fragmentation which causes congestion consuming bandwidth and CPU resources, lot of control overhead where for example, ICMPv4 has too many options, inefficient routing as a result of fragmentation and uncontrolled address assignment and minimal QoS support. The remaining part of this paper is organized as follows: Section 2 describes on the QoS provisions in IPv6, Section 3 discusses on RSVP, Section 4 elaborates on QoS architectures, Section 5 is simulation on queuing schemes, Section 6 is a new queuing scheme proposal and Section 7 is the conclusion.

## II. QOS PROVISIONS IN IPV6 PROTOCOL

Some of the QoS specific service elements of IPv6 in the Base header and Extension header can be explained as follows. *Flow* is defined when a series of packets require special treatment by the intermediate routers. The kind of treatment can be conveyed through the Base header of Extension header or through a corresponding control protocol. There can be multiple flows that are active from a source to destination. Other traffic may not be associated with any flows. The flow label would be non-zero for valid flow identification and would be zero, if it is not associated with any flow. If the packets belonging to a flow have Hop-By-Hop Extension header or Routing Extension header, except for the Next header field they all would be same in terms of contents. A violation of this rule can generate error messages from routers or receivers. *Flow Label field* is a 20 bit IPv6 header field that can be used by source to mark it for special handling. The flow label value can range from

00001 to FFFFF and can be chosen randomly which a router can look up to find the state associated with a flow.

In the original IPv6 (1996) proposal, the *Priority field* is a 4 bit IPv6 header field that helps the source identifies the desired priority of the packets. There are two ranges for priority as follows: 0 – 7 for fixed priority and 8 – 15 for real time traffic drop priorities. The first range is used in relation to congestion control and digits 0 – 7 corresponds to interactive traffic, attended transfers, unattended transfers etc. The second range does not generally back off in response to congestion. The lowest priority (8) in this range would be associated with packets that would be dropped first in congestion scenario and packets with highest priority (15) would be those that sender would want to drop the least. Later in 1997, the IPv6 header was redesigned to include an 8-bit Class field (when the flow label field was shortened from the initial 24 bits to 20 bits). This Class field also contained a 'D' bit for marking delay-sensitive traffic and three bits for network-wide priorities to allow for eight different traffic classes. The remaining four bits were reserved for possible use within a congestion-control protocol. In 1998, yet another redesign was proposed which replaced Class Header field with the Differentiated Services (DS) field. This is intended to supersede both the IPv4 TOS field and IPv6 Traffic Class field. The DS field allows a scalable service differentiation without the need for a per flow analysis at very router or end systems. Each DS compliant router contains a set of routines to deal with DS requests in relation to performance (to have more bandwidth or maximum bandwidth) or relative performance (identification and differentiation of service classes). The first six bits of the DS field are called as DS codepoint (DSCP) to register a per-hop behavior (PHB) that a packet requires at each intermediate router, where as the remaining two bits are unused. That makes the total number of unique DSCP combinations to be 64. A first pool of 32 codepoints is assigned for standard use, a second pool of 16 more codepoints is reserved for experimental or local use and the last pool of 16 codepoints is available initially for experiment or local use, but needs to be used as an overflow pool, if the first pool is used up.

There are two IPv6 extension headers that can be used to signal QoS requirements. They are Routing Extension header and Hop-By-Hop Extension header. The *Routing Extension header* can be used to request a specific route by indicating the series or sequence of IP address of the intermediate nodes to be used and this requires that the requesting node to have knowledge about the preferred route. The *Hop-By-Hop Extension header* is used to communicate a maximum of one router alert message per IP packet to every intermediate router in the path, so that routers can process them fast without changing the options. A value of 0 indicates that the IP packet contains a Multicast Listener Discovery message; a value of 1 indicates that the IP packet contains an RSVP message and a value of 2 indicates that the IP packet contains an active network message. Values from 3 to 65535 are reserved for future use [1].

### III. RESOURCE RESERVATION PROTOCOL

RSVP (RFC 2205) was invented as a signaling protocol for applications to reserve resources [2]. The signaling process is illustrated in Figure 1. The sender sends a PATH Message upstream to the receiver specifying the characteristics of the traffic. Every intermediate router along the path forwards the PATH Message to the next hop determined by the routing protocol. Upon receiving a PATH Message, the receiver responds with a RESV Message to request resources for the flow. Every intermediate router along the path can reject or accept the request of the RESV Message. If the request is rejected, the router will send an error message to the receiver, and the signaling process will terminate. If the request is accepted, link bandwidth and buffer space are allocated for the flow and the related flow state information will be installed in the router. RSVP has been modified since its inception and extended in several ways to reserve resources for aggregation of flows, to set up Explicit Routes (ERs) with QoS requirement, and to do some other signaling tasks [3].

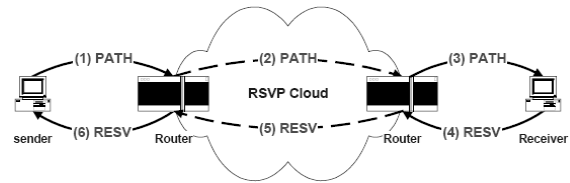

Figure 1. RSVP Signaling [3]

### IV. THE QOS ARCHITECTURES

There are two different QoS architectures that exists ─ Integrated Service and Differentiated Service. Both these schemes are based on traffic policing approaches that can even be integrated to allow for optimized QoS performance.

*A. Integrated Services Architecture*

The Integrated Services Architecture (IntServ) [4], [5], [6] is an approach where bandwidth and the needed resources are reserved per flow on end-to-end basis. This approach is very similar to establishing a telephone conversation by two communicating parties. During that conversation period the line is exclusively reserved for the parties engaged and cannot be sued by others. It allows the senders and receivers to make resource reservation, through exchange of explicit signaling messages, establish communication path through the network and allow packet classification and forwarding status information on each intermediate router along the path [1]. RSVP is specially designed to support IntServ. It reserves a portion of the output link in each router along the path of a flow. The sender periodically sends out a PATH message, which describes the type of traffic being sent and the resource requirements necessary to support the traffic stream. A receiver that gets the PATH message responds by sending a reserve (RESV) message toward the sender, tracing back through the same set of routers traversed by the original PATH message. At each router along the path, the RESV message is processed and the reservation is incorporated into the router (assuming the resources are available to honor the request). When the RESV message reaches the sender, an end-to-end reservation is established. Routers can reduce the QoS parameters of the RESV message to some degree if the reservation cannot otherwise be honored.

*1) Advantages of the IntServ Model:* This model has indeed a number of benefits. Firstly, the absolute service guarantees is promised in this design. The model allows RSVP clients to specify each service category in detail. Each flow can also be monitored to prevent it from consuming more resources than it had requested and reserved as RSVP runs in each router from the source to the destination. Because RSVP runs in each router from the source to the destination, each flow can also be monitored to prevent it from consuming more resources than it had requested, reserved, and presumably purchased. Another advantage of using RSVP is that it can use the existing routing protocols to determine the path of the flow between the source and the destination. By periodically retransmitting the PATH and RESV messages, the protocol can react to changes in the network topology. Just as these PATH and RESV refreshes can be used to change the path of a reserved flow, the absence of these messages can also be used to detect the loss of either the sender or receiver. When a router detects this loss, it de-allocates the resources associated with the reservation. One of the original goals of the IntServ group was to make QoS work for flows from one source to one destination (unicast) and from one source to multiple (potentially many) destinations (multicast). The RSVP protocol was designed to allow PATH messages to identify all the endpoints of a multicast flow and send the PATH message to each receiver. It also allows the RESV messages from each receiver to be combined into a single request at points in the network where a multicast flow would send the same flow on two separate links.

*2) Disadvantages of the IntServ Model:* Since the model is quite ambitious in nature, it carries with it some weaknesses too. Each of the routers should maintain a considerable amount of processing overhead and memory as all the routers participate in the resource reservation process. This model may not be that practical for short-lived flows, since the overhead of the reservation procedure is greater than the processing of all the packets in the packet flow. In situations where some modest level of QoS is important to short-lived flows, the IntServ model is overdoing it. The IntServ model also requires a substantial amount of state to maintain the reservation in a router and thus big buffers are needed. This state includes information to identify the specific flow, police excess traffic beyond the reservation made, track the flow's resource consumption history and schedule the traffic based on the reservation commitments made. The core of the network could contain many reservations (say, in terms of millions or more) that need to be controlled. Further, if a topology change occurs, all the reservations would need to be renegotiated simultaneously. Another issue that we need to think of with QoS is the economic aspects of quality. Specifically, within the service level of "best effort," all traffic is supposed to be treated equally and billed equally. When we introduce the concept of differences in service, we must also consider the billing model for these differences [6].

*B. Differentiated Services Architecture*

Differentiated Services Architecture (DiffServ) which is RFC 2475 [6], [7] is an innovative approach where the relative priority and type-of-service markings with the DSCP byte in the QoS sensitive packets are considered for dynamic reservation of resources. Explicit path setup and reservation signaling which was discussed under IntServ is not considered here. Hence this approach is not an end-to-end service and is based on the determination of PHB in each router. Here as stated before, the DS field in the packet is employed to indicate the QoS requirements and then used by DiffServ compatible routers to determine the further forwarding treatment with the packet. Classifiers and traffic conditioners can be used to select the needed packets for special treatment. This architecture covers a number of areas. The packets belonging to the aggregation must be identified, before traffic is grouped together into an aggregation. To ensure the service guarantees of individual aggregation, it is good to enforce the limits on the amount of traffic that any given user can inject. One part of the DiffServ architecture looks into these aspects, called classifying and policing. Within the domain of the DiffServ architecture, an intermediate router must be able to measure, shape, and drop packets in a flow. Another section of the architecture describes the difference between the edges and the core of the DiffServ network and how the classifying and policing mechanisms on the edge differ from those in the core. Some of the terms that we need to know in relation to DiffServ approach are as follows:

*Classifying:* A router must be able to collect the packets and recognize the flow to which it belongs to. On way to uniquely identify the flow is to note the source and destination IP addresses and, possibly, the source and destination port numbers, say for user data protocol (UDP) and transmission control protocol (TCP). This procedure of flow identification is commonly referred to as classification. Based on the DS field, the aggregation to which a packet belongs to can also be identified.

*Metering:* The measuring of the resource utilization or consumption, after a flow is classified is essential. To determine that the flow is not exceeding the agreed resource consumption limits, a router must first measure a flow's volume over some period of time. However, measuring flow rate is not sufficient to determine a flow's compliance. Sending higher rates of traffic at intervals is also referred to as sending bursts; when a host sends bursts, the resulting flow is said to be bursty. A typical QoS agreement often defines limits on the size of bursts as well as the maximum bandwidth for a flow. Consequently, a router must measure the flow rate as well as the size of traffic bursts and is often referred to as metering.

*Shaping:* A router can choose to handle and process a burst of packets, when a flow contains a burst of packets. One way is to process it normally if it falls within some predefined range. Another way is to absorb the burst and pace the packets out over a longer period of time. This pacing of the burst smoothes or even eliminates it altogether. The last alternative is to drop the packets in the burst that exceed a particular threshold. This threshold could be an upper limit on the number of packets in the burst that the router will have to retain at any point in time. Another threshold could be an expiry time, which is the amount of time that the packets would be kept before the data becomes stale. The procedure of holding the bursts and pacing the traffic is called shaping.

*Dropping:* Packets can be dropped or discarded by a router when a flow exceeds the negotiated rate or a burst exceeds a maximum threshold. To control congestion, dropping packets is a very common practice in any network scenario.

*Policing:* Metering, shaping, and dropping are all collectively referred to as policing.

One of the main features of the DiffServ scheme is its distinction between the edge and the core (middle) of an administrative domain. IntServ performs classification and policing on all packets matching a reservation in every router along the path. In contrast, DiffServ pushes most of the classification and policing functions to the edges of the administrative domain and simplifies the forwarding functions in the core of the domain.

*1) Advantages of the DiffServ Model:* This strategy addresses the scaling concerns by reducing all traffic to some number of traffic aggregations, each with a different set of QoS requirements. Unlike RSVP, no QoS requirements are exchanged between the source and the destination (commonly referred to as signaling), eliminating the inherent setup costs associated with RSVP. Short-lived flows benefit from DiffServ because the absence of QoS setup costs improves responsiveness and reduces the overhead for a quick discussion (a short-lived flow) with another host [4].

*2) Disadvantages of the DiffServ Model:* On the downside, the traffic aggregation model used by DiffServ adds some measure of unpredictability. Without reservation or signaling mechanisms and the traffic shaping that accompanies them, network traffic levels can become very dynamic. At arbitrary points in time, one part of the network may receive more traffic than another part. These variations in traffic volume and congestion would also occur in a single traffic aggregation, based on the time of day, the destination, and the bandwidth consumed by each flow in the aggregation. The ability to guarantee a particular level of service then becomes much more difficult. For this reason, the DiffServ model does not attempt to guarantee a level of service, but rather strives for a relative ordering of aggregations such that one traffic aggregation will receive better or worse treatment relative to other aggregations, based on the behavioral rules of each aggregation [6].

*C. IntServ/DiffServ Interoperation*

The interoperation of the two approaches seems to be a promising solution to provide end-to-end QoS in a scalable way. The basic idea is to use the DiffServ approach in the core network and the RSVP/IntServ in the access network. In this scenario, a key-role is played by inter working devices, called Edge Devices (ED), placed at the borders between these domains. See figure 2. In general, the Integrated Services over Differentiated Services approach end-to-end resource management is implemented by using IntServ model end to end across a network that contains one or more DiffServ region. The DiffServ is treated as a link layer medium of communication in such a case (say like LAN).

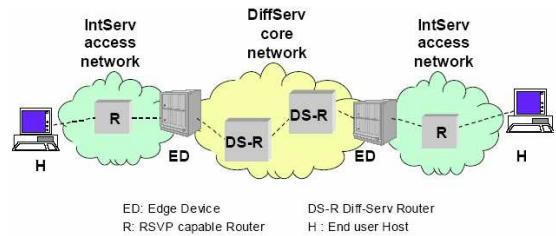

Fig 2. IntServ/DiffServ Interoperation Scenario [8]

Mapping is done at the boundary of the network, where IntServ requests are mapped onto the underlying functions of DiffServ network. Some of the aspects of mapping that should be considered are as follows [9]:

- To select the right PHB (per-hop behavior) or set of PHBs for the services requested.

- To perform the right kind of policing at the edge of the DiffServ network.

- To export IntServ parameters from the network that is DiffServ based.

This framework creates a two-tier resource allocation model where the DiffServ gives out aggregate resources in the core networks and IntServ allocates further the resources needed to individual users or specific flows on an as-needed basis. A policy server can be used to apply administrative policies to reservation requests. Based on the pre-defined rules, the policy server can decide how the IntServ requests be mapped to DiffServ model. The admission control policy allows critical packets to get priority to the assured service bandwidth over other traffic. Other complicated policies like who can use which forwarding classes and when may also be implemented [9] - [11].

V. BASIC SIMULATION FOR QUEUING SCHEMES

We did some basic simulation using three scenarios and got the results for the Queuing Schemes like *FIFO* (Fist in First Out), *PQ* (Priority Queue) and *WFQ* (Weighted Fair Queue) in relation to achieving QoS when doing video and voice traffic. This could help us to understand *IntServ* and *DiffServ* better. Later we'll discuss on *CQ* (Custom Queuing), *CQ with LLQ* (Low Latency Queue) and *WFQ with LLQ*. In FIFO, the first packet that arrives would be the first packet to be transmitted. The router would drop an incoming packet if the allocated buffer space in that router is full. PQ is a slight variation of FIFO queuing. Here each packet is marked with a priority by enabling some field. We enabled the ToS (Type of

Service) field in the IP datagram header. The router would then implement multiple FIFO queues, one for each priority class. Within each priority, the packets would work like in a FIFO queue. Thus high priority packets can get preference to get to the front of the queue. In WFQ, weight is assigned to each flow (queue) and each flow would be serviced in a round robin manner. The weight effectively controls the percentage of link's bandwidth each flow would get. Again ToS bit is used to identify the weight. It is algorithm that schedules low-volume traffic first, while letting high-volume traffic share the remaining bandwidth. This is handled by assigning a weight to each flow, where lower weights are the first to be serviced.

In the first scenario, two subnets were connected through 2 routers with a 56Kbps PPP central link. The central link between subnets is a bottleneck. One subnet is for servers and the other subnet is for clients. The test network created had options for voice, video and data communications between the central link, when client accesses the servers. Firstly we checked the packets dropped in FIFO, PQ and WFQ.

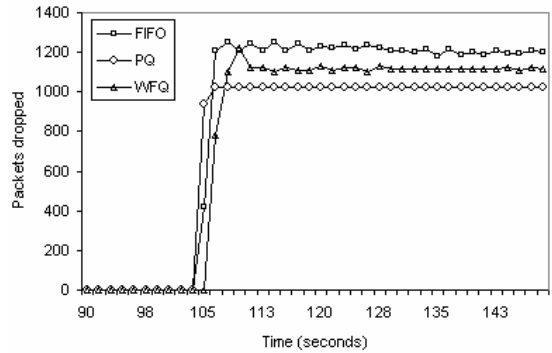

Fig 3. Packets dropped when using Queuing Schemes like FIFO, PQ and WFQ.

Figure 3 shows that the packet drop is relatively high for FIFO and WFQ and PQ performs better. Next we checked on the video and audio packets with the different queuing schemes.

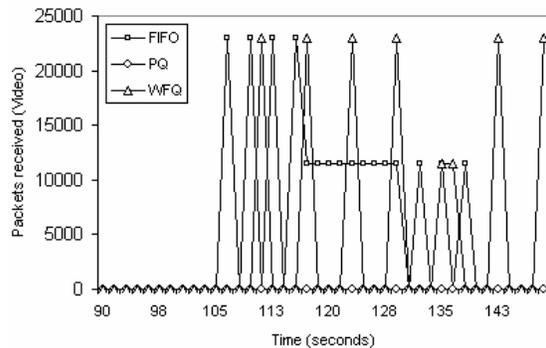

Fig 4. Video packets received when using Queuing Schemes like FIFO, PQ and WFQ.

Figure 4 shows that the video packets received is relatively better for WFQ, compared to FIFO which is doing better than PQ that is close to x-axis. Figure 5 for voice traffic shows that PQ scheme is working well in relation to others, followed by WFQ and FIFO.

In the second scenario, RSVP nodes for voice communication were used and added to scenario one (stated above) and simulation was performed on them with 1.544Mbps central link between subnets. Comparison was done with and without RSVP as in figure 6.

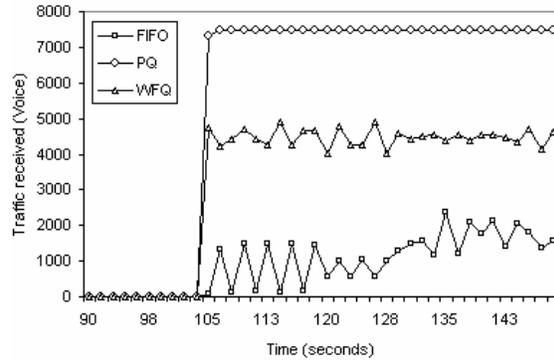

Fig 5. Audio packets received when using Queuing Schemes like FIFO, PQ and WFQ.

In Figure 6, the delay for RSVP voice communication is lower and shows that the bandwidth of the backbone of the central link can play a major part in keeping the delay lower.

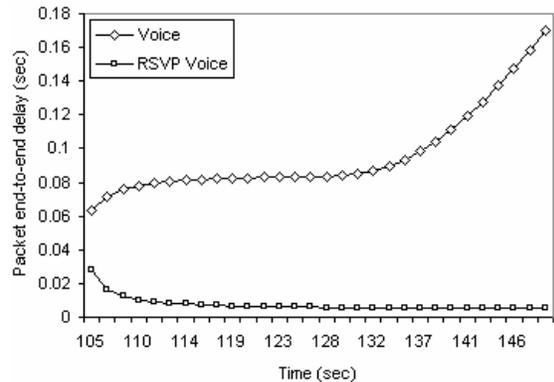

Fig 6. The time average view of packet end-to-end delay is shown for RSVP Voice vs. Voice Communication over a 1.544Mbps central link between subnets.

In the third scenario, we used a network with four video clients accessing four servers through two routers in between. The four stations on each subnet are connected to switches which in turn are connected to routers. The link between the routers is a potential bottle neck with a low bandwidth of 1 Mbps. Six scenarios were created for FIFO, PQ, CQ (Custom Queuing), CQ with LLQ, WFQ and WFQ with LLQ and the delay graph is shown in figure 7. LLQ is a low latency queue which acts as a absolute priority queue within CQ or WFQ and no other queues can be serviced until LLQ is empty. In CQ, ToS field is used to differentiate traffic from different queues and works in a round robin manner. ToS field is also used in WFQ to associate weightage. The delay was found to be worst for FIFO, followed by WFQ and WFQ with LLQ. PQ and CQ approaches worked fine. That shows us that WFQ can thus give high delays when the link bottle neck is a factor. That leads us to propose a better queuing algorithm that relates with a modified version of WFQ.

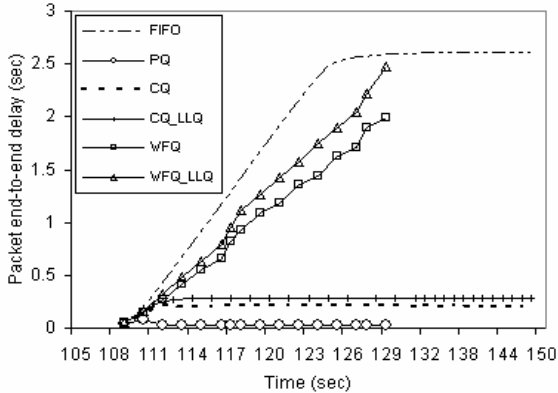

Fig 7. The packet end-to-end delay is shown for six scenarios for video traffic over a low bandwidth central link between subnets.

## VI. PRIORITIZED WFQ WITH ROUND ROBIN PROPOSAL

We propose a new queuing algorithm (Prioritized WFQ with RR) that is a modified version of Weighted Fair Queuing with Priority queuing and round robin scheme implemented on separate queues. Flow label and Priority fields (or its modified version) in the IPv6 header can be used to classify packets based on flow and priority [11]. The steps can be as follows:

1. Allow the traffic of incoming packets to router. The classification of the packets can be, for example – ToS based, protocol based or port based and weight is assigned to each incoming flow.
2. The packets with highest weight $w_1$ are channeled into a Priority Queue 1 ($PQ_1$). Here packets coming in are marked for priority and sub-queues ($PQ_{11}$, $PQ_{22}$ etc.) are maintained for each priority. Within each sub-priority queue, packets are managed in a FIFO manner.
3. Other packets with next highest weight $w_2$ are maintained in Priority Queue 2 ($PQ_2$) and a set of sub-priority queues ($PQ_{21}$, $PQ_{22}$ etc.) are maintained as explained in step 3.
4. When the router services these queues, it starts with the Priority Queue 1 and so on.
5. The whole scheme then works in a round robin manner, where each queue is given a time slice and where Priority Queue 1 gets more share of time $t_1$ and Priority Queue 2 gets the next highest time $t_2$ and so on.

If there are 3 general queues, the bandwidth for each queue $PQ_1$, $PQ_2$ and $PQ_3$ would be allocated as: $BW_1 = w_1/(w_1+w_2+w_3)$, $BW_2 = w_2/(w_1+w_2+w_3)$, $BW_3 = w_3/(w_1+w_2+w_3)$ respectively, where $BW_1 + BW_2 + BW_3 = 100$. The sub-queues associated with $PQ_1$ would occupy bandwidth based on the priority assigned. For example, within $PQ_1$ if four priorities are assigned as $p_{11}$, $p_{12}$, $p_{13}$ and $p_{14}$, the bandwidth for each of the 4 sub-queues would be allocated as: $BW_{11}=p_{11}/(p_{11}+p_{12}+p_{13}+p_{14})$ where $BW_{11}$ is the bandwidth of first sub-queue within $PQ_1$, $BW_{12}=p_{12}/(p_{11}+p_{12}+p_{13}+p_{14})$, $BW_{13}= p_{13}/(p_{11}+p_{12}+p_{13}+p_{14})$, $BW_{14}= p_{14}/(p_{11}+p_{12}+p_{13}+p_{14})$ respectively, where $BW_{11} + BW_{12} + BW_{13} + BW_{14} = BW_1$. The same applies to other sub-queues under $PQ_2$ and $PQ_3$. The above scheme helps to give a fair treatment to all packets, with the highest priority packets or packets with high flow ranking getting the preferred treatment while not starving the lowest priority packets. The choosing of right or optimal weight $w_i$, priority $p_{ij}$ and time $t_i$ is important and that ensures the efficiency of proposal.

## VII. CONCLUSION

This paper presents a survey on the Quality of Service options with IPv6 networks and some basic simulation on different queuing schemes. The native support for QoS in IPv6 is highlighted. It then discusses various QoS architectures with its pros and cons. A combined architecture of IntServ and DiffServ is discussed. We also propose a new queuing scheme (Prioritized WFQ with RR).